\documentclass{article}
\usepackage{amsmath, amsthm, amssymb, mathtools, outlines}
\usepackage{tikz, physics}
\usepackage{array, float} 

\usepackage[colorlinks=true, linkcolor=blue]{hyperref}
\usepackage{graphicx, authblk} 
\usepackage[english]{babel}
\usepackage{natbib}

\allowdisplaybreaks
\numberwithin{equation}{section}

\title{Intersubjective Agreement about Measurement Outcomes Is Unnecessary in QBism}
\author{Gino Elia, Jennifer Carter, and Robert Crease}
\affil{gino.elia@stonybrook.edu, jennifer.carter@stonybrook.edu, robert.crease@stonybrook.edu}
\affil{Stony Brook University, Philosophy}
\affil{Stony Brook, New York, United States}

\date{\today}

\begin{document}

\maketitle

\begin{abstract}
    The thought experiment called ``Wigner's Friend" has experienced a renewal of interest for interrogating the meaning of intersubjectivity and objectivity in quantum mechanics. These new inquiries extend to investigations at the intersection of phenomenology and QBism. Philosopher of physics Steven French argues that QBism does not give assurances that Wigner and friend must agree on the same quantum state or measurement outcomes. In this article, we draw on Wigner's Friend to argue that an external guarantee for agreement on either quantum states or measurement outcomes is unnecessary. We defend the view that the quantum formalism is already inherently intersubjective in the way required to sustain objectivity. Here we explore the QBist notion of reciprocity, which treats Wigner and friend as physical systems taking mutual actions on each other. The QBist notion of reciprocity leads to a sharper characterization of what it means to objectify quantum systems with the formalism. Drawing on phenomenological resources, we argue that state assignments for quantum systems, including those for Wigner and friend, are a form of objectification. To assign a quantum state is to objectify a phenomenon as a quantum system, to treat something as the sort of object to which the formalism applies. Our argument accounts for why the quantum formalism does not radically change in application for different systems because the systems themselves exceed their formalization.\\

\end{abstract}


\section{Introduction}

In \textit{A Phenomenological Approach to Quantum Mechanics: Cutting the Chain of Correlations}, Steven French argues that Fritz London and Edmond Bauer already offered a phenomenological approach to quantum mechanics in 1939 \cite{french2023}. In \textit{The Theory of Observation in Quantum Mechanics} (1939), London and Bauer (LB) rejected an interactionist account of quantum measurements, where the interaction between an observer and a quantum system induces ``wave collapse” \cite{london1982}. LB characterized measurement with a more phenomenological approach, where an act of observation distinguishes the observed outcome and the observer as poles of the same relation. They described how quantum states represent our knowledge about a possible measurement in which the act of observation produces an objective measurement outcome. When we update our quantum state, we distinguish ourselves from the rest of the system.\\

The main confusion French addresses in the LB account is that consciousness plays a causal role in such an event. LB's pamphlet on measurement is often regarded as a summary of von Neumann’s position in \textit{Mathematical Foundations of Quantum Mechanics} (1932) \cite[227-232]{vonNeumann1955}. There, von Neumann distinguished between the continuous process of the time evolution of the wave function and the discontinuous process of measurement. In the latter process von Neumann discusses how the cut between the observer and the observed is largely arbitrary, including how the role of an observer in a physical description need not be identified with their physical body \cite[272]{vonNeumann1955}. The von Neumann account was widely interpreted as saying that consciousness played a causal role in measurement. In the 1960s, as French points out, the position of LB came to be erroneously identified with the interpretation of von Neumann’s, effacing the value of the LB account in the process \cite{MargenauWigner1997, Putnam1964}. The first half of French’s book explains the origin of the historically inherited misconception to articulate the LB account as well as its phenomenological motivation.\\

In this article, we do not question most of French’s analysis of London and Bauer’s work. Our aim is to contribute to a phenomenology of quantum mechanics that challenges the view that intersubjective agreement about measurement outcomes is necessary for objectivity. Do we require a guarantee that observers will agree on their quantum states and measurement outcomes? Is it a problem if we do not know if our communications will inevitably converge? The worry is that, without a guarantee, making quantum states or measurement outcomes personal cannot support or explain intersubjective agreement.\\

In what follows, we offer a series of heuristics to answer these questions using the Wigner’s Friend thought experiment. We argue that French’s worry about intersubjectivity is adequately addressed by distinguishing between intersubjective agreement about states/outcomes and intersubjective agreement about how to use the quantum formalism. The intersubjective demand for mutual agreement on how to use the formalism is called ``reciprocity." Reciprocity lives higher upstream in the epistemic pipeline than agreement on states/outcomes. It stipulates that Wigner and friend agree on how to structure their probabilities with quantum theory, but quantum measurements do not in themselves require intersubjective agreement,\\

We acknowledge that the term ``intersubjectivity" fractures into multiple competing demands, although the distinction between an account of intersubjectivity and intersubjective agreement, while common in philosophy of science, does not seem to have been as widely adopted in quantum foundations literature. Reciprocal use of the formalism is an intersubjective demand. Intersubjective agreement on states/outcomes is also an intersubjective demand. Nevertheless, in a Wigner’s Friend scenario, our reading is that we do not get both, and that tells us something important about intersubjectivity in quantum mechanics. Absent a global set of shared facts prior to one's measurements, the demand for convergence can only be satisfied once each has communicated one's results.\\

We distinguish between reciprocity in quantum mechanics and one applicable to QBism. The former describes how the formalism emphasizes the positionality of observers when structuring their probabilities. The latter emphasizes how Wigner and friend are independent agents whose measurement outcomes are personal. The QBist notion leads to an altogether novel picture of our experience of the world. In both cases, reciprocity sits better with a long history of quantum mechanics telling us not to go looking for objectivity at the level of individual events. Further, we find that the LB account does not require a guarantee about intersubjective agreement either, indicating that their remarks on intersubjectivity are closer to QBism than French suggests.\\

The topic of reciprocity leads to a more pressing question for the phenomenology of quantum mechanics. How should we understand the QBist prohibition that one cannot assign a quantum state to their own person? A generic account of intersubjective agreement may suppose the existence of a global set of facts on which Wigner and friend agree, but this also leads to the idea that Wigner’s and friend’s quantum statistics are interdependent. Someone is right; the other is wrong. True intersubjectivity, in the QBist view, requires that both Wigner and friend rely on the formalism as it stands. This leads to the view that each agent wields the formalism independently. If quantum mechanics is a single-user theory, one cannot assign a quantum state to their own person because their predictions then depend on their own actions.\\

In our view, QBist concerns about self-reference have more to do with the question of objectification: what does it mean for an agent to ``objectify” a quantum system with their quantum state? What does it mean for Wigner to assign a quantum state to the friend? For most accounts, Wigner’s pure state appropriates the friend’s subjectivity, leading to a Schrödinger’s cat paradox. For QBism, quantum state assignments are personal for Wigner and for the friend relative to their own positions. Quantum states do not pierce into a quantum system to figure out ``what’s really going on” inside \cite[194]{french2023}.\\

In response to these questions, we suggest that objectification in phenomenology is similar in spirit to QBism's take on the formalism. Objectification does not mean rendering real-world phenomena into objects. It is a complex, creative act for a subject who ``thematizes” phenomena in an interested or purposeful way. This provides insights into why QBism regards Wigner and friend as quantum systems taking mutual actions on each other, why QBism also regards ``quantum systems” as being more than just quantum systems, and why QBism forbids self-reference with a quantum state. These three ideas are related through the notion that assigning a quantum state to a system is precisely not to fully objectify it. Objectifying acts supply material for a description of nature but have no ontic hold on the world. Assigning a quantum state is inherently selective according to the relevant beliefs we have about the quantum system and what experiences we expect to occur.\\

We do not intend this position on objectification as a criticism of QBism. Rather, QBism has a distinct flavor of what is called the ``universality" assumption of the quantum formalism \cite[7]{schmid2024}. Whatever one wishes to say about quantum states, we should understand it as applying to all physical systems in the same way. If it is not coherent to fully objectify oneself, it should be for the same reason that one cannot do so with any other physical system. This point is sometimes misunderstood in phenomenology. Phenomenologists often emphasize that the subject cannot be fully objectified as an object, but this is not because the subject/agent has the secret ingredient of sentience. The idea is simply that to objectify any bit of experience is to delimit it as an object of study, not to reify it. The phenomenological objection to reification does not discriminate what sort of experience we are talking about.\\

In section 2, we narrate the basic setup of Wigner’s Friend to motivate various philosophical conflicts surrounding intersubjectivity. In section 3, we review the relevant features of QBism and their corresponding resolution of Wigner's Friend. In section 4, we develop the concept of intersubjectivity in terms of reciprocity in quantum mechanics and QBism. We offer reciprocity as a solution to French's concerns about intersubjective agreement. In section 5, we turn to our reading of the LB account to further reduce worries about intersubjectivity. In section 6, We address specific critiques from Steven French and also Emily Adlam against intersubjectivity in QBism. We attempt to explain their concerns about intersubjectivity stem from elsewhere than QBism, and that appeals to additional assumptions such as absoluteness of events or tracking observer outcomes does not resolve them. In our view, the sort of intersubjective agreement desired is a ``metaphysical fifth wheel" that evades the physical reasons one would cite to doubt the friend's report of her result. However, if one insists on absoluteness, one which could serve as a guarantee, then this assumption has practical implications for Wigner's Friend. In section 7, we put forward a QBist response that one cannot demand a guarantee for intersubjective agreement and withhold a possible measurement scheme to preview the objective outcome independently of Wigner and friend. However, such a measurement scheme would appear to violate the rules of quantum mechanics. In section 8, we offer our own contribution to what we believe is the more urgent question - quantum universality - by discussing the phenomenology of objectification. Our entrance into the subject is again through the QBist notion of reciprocity and the possibility of self-reference with quantum states. We argue that universality implies that all quantum systems be treated the same way, for which phenomenological accounts of objectification offer a resolution.\\

To flag a common theme, the meta-criteria for agreement and communication are tacitly appealing to reasons not found in physics, but metaphysics. As such, our strategy is not to ``solve" but rather disarm them. For us, one hint that we should attempt dissolution, before resolution, is the haunting similarity between intersubjective agreement in Wigner's Friend and perennial worries over skepticism in philosophy. As with skepticism, it is reasonable to ask for a practical example that supports the skeptic's claim that without a metaphysical guarantee aside from the experiment, one cannot have knowledge of the world, of others, or of their measurement outcomes. 

\section{Wigner's Friend}

In this section, we offer a qualitative description of Wigner's Friend for the purpose of showing that the essential features of the puzzle are already present without the technical details of quantum mechanics.\\

The Wigner's Friend thought experiment originated in 1961 in Eugene Wigner's essay, “Remarks on the Mind-Body Question.”\footnote{ See E. P. Wigner, “Remarks on the Mind-Body Question," in The Scientist Speculates, edited by I. J. Good (William Heinemann, Ltd., London, 1961), pp. 284-302; reprinted in E. P. Wigner, Symmetries and Reflections: Scientific Essays of Eugene Wigner, (Ox Bow Press, Woodbridge, CT, 1979), pp. 171-184.} Wigner and his friend, Alice, make an agreement on a set of procedures for an experiment. Alice will enter and seal the laboratory while Wigner remains on the outside. At a prearranged time, Alice will perform any measurement she likes. The measurement in the original thought experiment had two possible outcomes – “flash”  or “no flash” – with equal probability, where the probabilities for the outcomes are calculated in the usual way with the Born Rule. Afterward, the lab is opened and Alice tells Wigner what the outcome was.\\ 

From Wigner’s perspective, there is no measurement outcome until Alice tells him her result. From Alice’s perspective, assuming she agreed to follow through with the experiment, she performs the measurement and sees either “flash” or “no flash.” It is intuitive for Alice to also predict that Wigner will see the same result as her after the lab is opened.\\

Wigner does not have to passively wait for Alice to open the lab; he can make his own prediction. However, assuming Alice sticks to the agreement to make her measurement at the appropriate time, Wigner cannot predict that he would see a flash (or no flash). He can only predict that his friend may see it. That is, Wigner’s prediction is tied to his expectations for what the friend will see and do in the lab. Wigner’s prediction is therefore not about the flash itself, but the friend-flash system.\footnote{Renner and others have suggested on multiple occasions that Wigner's prediction could absorb the distinction between representations and meta-representations (what Wigner believes about the predictions of Alice) \cite{frauchiger2018, hausmann2025}. If so, experimenters would be able to adopt the predictions of others as their own. In our opinion, we do not see how this is possible. The distinction between representations and meta-representations is not a matter of probability. The positionality of each physicist plays an essential role in the state assignments.}\\

Do Wigner's and Alice's predictions match? It is clear that they do not, since their systems are different. Yet, a more important question is, does quantum mechanics itself discriminate between these two predictions? It does. Quantum mechanics inclines Wigner to treat the lab contents, including the friend and her quantum system, as one combined system. He thus assigns an entangled state to the system that continuously evolves in time. This has two implications. First, we can intuitively understand Wigner assignment of a pure state to the friend-lab system since Wigner himself does not have access to Alice’s lab and is not doing the measurement. Wigner’s expectations are tied up with what he believes about Alice’s actions, including that she will stick to the “agreement,” that is, she will execute the plan to perform the experiment. All these circumstances are rolled into his pure state assignment.\\

The alleged contradiction occurs at the intermediate time when Alice performs the measurement and sees a measurement outcome, either “flash” or “no flash.” However, from Wigner’s point of view, the system is still “in” an entangled state. The puzzle arises insofar as one accepts that according to the quantum formalism, both Wigner and Alice are correct. Whose description is right?\\

The answer one gives to the puzzle depends on one’s interpretation of quantum mechanics. As Baumann and Brukner see it, no matter what we think about the interpretation of the Wigner’s Friend thought experiment, it exposes different intuitions about how we should apply quantum theory \cite{frauchiger2018, baumann2019, brukner2018, debrota2020a, del_santo2024, Healey2018, proietti2019, zukowski2024}. Trade-offs arise between the objectivity of the quantum state vs. the objectivity of the measurement outcome. On first pass, the most tempting resolution is to say that Alice's measurement outcome is objective, in which case the friend is right, and Wigner did not have the whole story. However, Wigner cannot simply back off the implications of his entangled state. There are setups that show the friend's predictions are \textit{incorrect} compared to Wigner's, that is, if he regards his quantum state as descriptive of an underlying system \cite{baumann2019}. Next, one might suggest that Wigner's quantum state is correct, and the friend's measurement outcome should not be regarded as objective, in which case, we must now explain why quantum mechanics vindicates the friend's initial state assignment. The conventional wisdom is that we should expect them to assign different quantum states, which heavily indicates each is not riveted to an actual state of affairs. The tension here occurs if one \textit{does} wish to maintain that a global state of affairs appeared when Alice performs her measurement \textit{and} that they are allowed to assign different states. It seems as though one is indicating a third state assignment, separate from both Wigner and friend, one that knows the measurement outcome really occurred. This suggests a possible measurement scheme that would reconcile their observations (a quantum ``spy cam" if you like). What is generally not an accepted resolution, however, is appealing to the inevitable agreement on the measurement outcome at the third time. Although Wigner and Alice may eventually share the same opinions on the result, such sharing does not explain how quantum mechanics gives, correctly, different predictions for Wigner and friend. Their having assigned different quantum states is no longer controversial in quantum foundations. What remains disputed is whether or not measurement outcomes are ``personal." It is ambiguous how to regard Alice's measurement outcome before the lab is opened. It is as if the measurement outcome was only ``for" Alice.\\

Of all the tensions the no-go theorems propose for Wigner's Friend, the most provocative is between locality (as well as ``local agency") and the absoluteness of events \cite{schmid2024, Bong2020}. Locality means no spooky action at a distance, whereas local agency is the assumption that measurement settings in an experiment are uncorrelated with events outside its causal influence. Absoluteness says that measurements are global events available to everyone when they occur. Out of a preference for locality due to special relativity, the momentum of opinion has currently shifted against the assumption that measurements are global events. Many have pointed out that the entire disagreement turns on the assumption that states concern the same measurement event \cite{del_santo2024, brukner2015, jones2024, pienaar2024}. Putatively, drop absoluteness; the debate dissolves.\\

Locality, absoluteness, and variations on these concepts have been defended as ``preconditions" for the possibility of science \cite{adlam2022, adlam2024, Howard1985}. Einstein took the position that we could not do objective science unless we could isolate some piece of the world for an experiment, sequestered from our other experiments \cite[187-188]{Howard1985}.\footnote{According to Don Howard, Einstein favored the independent existence of spatially separated systems, and ``locality," that a characteristic experiment on one system does not have an immediate effect on the other. Einstein's philosophy addresses the ability to isolate an experiment from external influences.} The notion of an experimental demonstration seems only to cohere if the procedure, and not Alice's actions, is counted as relevant to the conclusion. With local agency, what is at stake is discarding irrelevant information from the relevant experimental context. On the other hand, it seems obvious that we explain our agreements because a measurement outcome occurred in the world. An event, or an object inscribed with the physical memory of an event, should serve as a ground of appeal independently of our observations.\\

The usual defense of absoluteness is that something should explain the convergence of observer communications. French holds that we need some assurance that Alice’s measurement outcome is the same one that Wigner will see after the experiment \cite[175]{french2023}. Tracing its roots upstream, a ``fact of the matter” appeared when Alice made her measurement, but before the lab was opened. In this way, French’s call for assurances that the two will agree after the lab is opened harks back to the disagreement in their quantum state assignments. Wigner could sincerely believe that Alice saw an outcome, but if he does not have his own way of verifying this, he cannot update his pure entangled state to a mixed state. If they are ``fated to agree” on a measurement outcome, then this destiny should manifest itself in the quantum formalism. Could we make sense of objectivity if every event was messily correlated with everything else? Could we make sense of objectivity if the agreement between Alice and Wigner was a matter of coherence?\\

Many of these debates are still in their early stages. Often philosophers will rediscover anxieties over skepticism of the world (Descartes' Evil Genius), the problem of other minds (Chalmers' zombies), the convergence of truth (Charles Peirce's famous clash with William James), and the non-inferential ground of appeal (Sellars' Myth of the Given). The philosophical conflict over consensus in Wigner's Friend likely has some of the same aspects as general epistemic skepticism, but is buried in the formalism and takes conceptual digging to resurface in a familiar way. The dialectic of skepticism usually goes something like the following. I present evidence that some event \textit{A} occurred. The skeptic retorts: how does one know that the event \textit{A} I saw is the same event \textit{A} they saw or could see under a similar set of conditions? At this point the skeptic and the philosopher merge, for either the skeptic is dissatisfied with all the reasons one could offer as support for one's observations, or one requires a super-observational criterion - a metaphysical guarantee outside the relevant experimental context - that underwrites one's observations. For it will be argued that some extra criterion must be in place for evidence or reasons to count as such for event \textit{A}. Whatever one's position about skepticism, analogues in Wigner's Friend are forthcoming, such as the question: how does Wigner know that the measurement outcome conveyed by Alice is the \textit{same} as what Alice really saw? It would be surprising for this to be a problem uniquely forced on us by quantum mechanics, since nothing about the theory is necessary to articulate the skeptic's view. This is a clue that worries over skepticism and intersubjectivity may be red herrings. We formulate the following as a heuristic:\\

\textbf{The Chain of Evidential Reasoning}: evidence, justification, and possible doubts are grounded in the relevant physical context.\\

The chain of evidential reasoning (CER) conveys the belief that when we \textit{doubt} somebody's measurement outcome, we never appeal to ``the inevitability of convergence" or a meta-criterion. We go to the milieu of the relevant information and ask questions (e.g. was so-and-so telling the truth? Was the measurement device faulty? Did somebody flub the time to do the experiment?).\footnote{CER is partly inspired by Cavell's advice: ``I know no more about the application of a word or concept than the explanation I can give, so that no universal or definition would, as it were, \textit{represent} my knowledge - once we see all this, the idea of a universal no longer has its obvious appeal, it no longer carries a \textit{sense} of explaining something profound" \cite[188]{cavell1979}.} The skeptic's plight is possible after we have exhausted CER, in which case, we believe it is reasonable to ask, is the skeptical objection a difference that makes a difference? Note that CER is not an outright rejection of modalities, or an analysis of presuppositions \cite[262]{French2014}. Rather, it is a heuristic that is supposed to help identify which modalities are relevant to theorize in the first place (and from which claims to elaborate them).\\

With CER in mind, we offer a few more reasons that locality is preferable to absoluteness. First, some version of locality appears to be taken on as a premise when setting up an experiment. It is not clear how CER would apply if one could not reliably isolate a set of reasons within a given context for why this or that experiment was successful. Second, it is often pointed out that locality is physically motivated from special relativity, albeit in a generalized form \cite{schmid2024, Howard1985}. Absoluteness, on the other hand, does not directly make an appearance in CER, but as a retro-causal account, nor is it directly motivated by any physical theory. One hopes we might preserve both, but if the no-go theorems invalidate one, we would think that losing absoluteness is no loss at all, since it is not a part of CER to begin with. Matters here are not helped when we flat out identify \textit{objectivity} with a metaphysical desideratum like absoluteness or locality. There is some reason to be suspicious when we flag a single criterion as being necessary for objectivity. One cannot easily get away with proposing an objective framework for doing science without possibly infringing on the ability for science to make claims. In the event that neither locality nor absoluteness prove essential, then we end up explaining the objectivity of evidence some other way (without philosophical anxiety!).\\

Regarding intersubjective agreement, we find the notion meaningful if it offers new insight into the phenomena under investigation. That is, we can ask why we agree. When we make this inquiry, we do so with reasons from our own experiences, such as Wigner and Alice both being able to inspect the device after the lab is opened. Attributing facts to a shared reality passes through the communications between Wigner and Alice. Using the formalism in the correct manner allows them to structure their probabilities in such a way that intersubjective agreement is possible. The preparation of the lab and procedures for the experiment set up the contours of agreement/disagreement for Wigner and friend, but their prior interactions do not compel a specific outcome. A genuine benefit of intersubjectivity is to seek convergence about experiences that could have been otherwise, not merely to rehash confirmation. The notion of intersubjective agreement would be vacuous if it only referred to observers stating the same outcome. It would also be vacuous if it only referred to an already presupposed outcome because the epistemic gain of intersubjective agreement would be null. If Alice already thought she could verify her outcome with certainty without anyone else, why double check anything? In neither case does a presupposition of absoluteness assist in interpreting experimental outcomes. With respect to intersubjectivity, Wigner’s Friend, after all is said and done, appears ordinary, not exceptional.\\

In the next section, we review the essential features of QBism for Wigner's Friend. While the absoluteness of observed events criterion has caused much debate in quantum foundations, the way in which quantum mechanics forces us to an austerity about being intellectually honest about what we say we observe is not new. Absoluteness is not the most crucial criterion about which to throw down the gauntlet. The assumption does no work for the Wigner’s Friend thought experiment and thus is a fifth wheel. The structure of Wigner’s Friend forces Wigner and Alice to be explicit about their positions and their measurements, but they could have done so anyway without quantum mechanics.

\section{QBism and Wigner's Friend}

The QBist response to Wigner’s Friend does not make use of absoluteness, meaning that Wigner and Alice are both right if they use the formalism correctly. The deeper sense of objectivity in the QBist approach rests with the formalism itself. For the QBists, this amounts to saying that Wigner’s prediction is \textit{for} Wigner; Alice’s prediction is \textit{for} Alice \cite{caves2007, fuchs2025}. The alleged contradiction dissolves because from Wigner’s perspective, he does not get a result until Alice tells him. Quoting DeBrota, et al.:
    \begin{quote}
        In QBism, the quantum formalism is only used by agents who stand \textit{within} the world; there is no God’s-eye view...Wigner treats his friend, the particle, and the laboratory surrounding her...as a physical system external to himself...the friend must reciprocally treat Wigner, the particle, and her surrounding laboratory...as a physical system external to herself. It matters not that the laboratory spatially surrounds the friend; it, like the rest of the universe, is external to her agency, and that is what counts \cite[3]{debrota2020a}.
    \end{quote}
The QBists offer a series of interlocking tenets for how to regard quantum theory. All five are required for the approach to be consistent \cite[4-6]{debrota2020a}:
    \begin{quote}
    1. A measurement is an agent's action on their external world whose consequences matter to the agent. Measurements are active interventions in the world, not passive records of something preexisting.\\
    
    2. The quantum formalism is an appendix to Bayesian probability theory and therefore pertains to an agent's decision making. A quantum state is then a statement of expectation regarding what consequences one expects to receive.\\
    
    3. The measurement outcome is personal. That is, the consequence of an action is for the agent and not publicly available to everyone.\\
    
    4. The quantum formalism is therefore single-user and normative. My use of the Born Rule, for example, does not prescribe how another agent should use it. The friend, for example, does not have to adopt Wigner's quantum state as their own, since their measurement is personal.\\
    
    5. Being statements of expectation, a quantum state of probability-one nevertheless expresses an agent's total confidence in the consequences of a measurement. However, a statement of belief, being personal to the agent, does not guarantee nature will produce a certain outcome.
    \end{quote}   
These propositions work together as a unit. Much of the QBist argument begins with the idea that quantum states are not ``things in nature” but probabilities for the consequences of taking actions on quantum systems. Suppose that we accept that quantum states are probability distributions for agents. Both Wigner and Alice must use the Born Rule to see how their state assignments result in different predictions. The Born Rule is an objective invariant for the theory which acts as a coherence condition for our probabilities. The QBist program reformulates the entirety of quantum theory as a kind of probability theory that reflects this idea. Quantum states, measurement operators, and unitaries are also subjective judgments — expressed as probability assignments — about quantum systems. If one’s probability $p$, one’s quantum state $\rho$, or one’s measurement operator $E$ does not satisfy the Born Rule, then nature is telling us that the set is inconsistent, but does not say which one should be revised \cite[5]{debrota2020b}.\\

More to the point, if a quantum state is interpreted as a Bayesian probability, then it follows that this state is not a state of the system. The QBists often use a lottery example to explain this point. My evaluation of a horse consists in a list of bets suggesting who I think will win or lose against the horse in a race. My bet is the quantum state. It makes no sense to view my bet as a property of the horses or of the race. A quantum state is not a state ``of” a system but an expectation of what happens when an agent takes an action on a system. Even if I feel ``certain” about some state of affairs, my certainty is still an expectation. There is no magical moment when probability stops being a set of beliefs, even at probability-one \cite{caves2007}. Even when we feel certainty that $(p = 1)$, we do not have a metaphysical promissory note that the event will occur.\\

The notion that probability-one does not imply that the outcome happens by necessity is likely the most controversial of the tenets. However, once we say quantum states are statements of expectation, not predicates of systems or of reality, it is not possible to evade the implication that probability-one still expresses belief. One can arrange conditions for certainty when preparing a measurement, just as one can bet on a horse race with only one horse. Even so, the bet would not categorically become “of” the horse just because there is only one. Upping a quantity to its maximum does not change the category of measure. The contingency of rain or of the horse getting sick could still jeopardize the race. A disgruntled bookie could say, ``In principle, if the race had happened, then the only horse would have won,” but we see this expression is now a bet on another bet. When Wigner and Alice take actions on their respective systems, the quantum formalism stipulates that they assign quantum states that account for expectations of their own experiences. In terms of argumentative structure, \textit{notice it is not the case that QBism actively drops absoluteness due to the idiosyncrasies of the Wigner’s Friend puzzle.} QBism simply does not have that assumption.\footnote{See Fuchs' discussion of personalist measurement outcomes in \textit{My Struggles with the Block Universe} (2015), 675-677 and 1010-1014.}\\

Here it is important to discuss whether QBism has ``two levels" of personalism or just one level consistently followed through \cite[2]{fuchs2025}. The way that Wigner's Friend seems like it motivates two levels is this: first, one asserts that the quantum formalism warrants Wigner and friend to assign different quantum states (level one). Next, when the measurement outcome occurs for the friend, but not necessarily Wigner, one seems inclined to give up the notion of a global event and instead amend ``the" measurement outcome with ``one's" measurement outcome (level two). These conclusions follow more naturally if one does not conceive of the gold standard of objectivity for \textit{individual events} as opposed to rules, laws, structures, normative prescriptions - all of which point to ontologies ``upstream" from ``this" or ``that" observation.\\

The question is how the two levels of personalism are connected. How much wiggle room is there to deny that measurement outcomes are personal once one has already agreed that quantum states are personal? If Wigner and friend are correct in their state assignments, but a new event occurs in the world, then at that moment, their information is ``incomplete." This seems to hint at a possible state, the ``true state," with a corresponding measurement scheme, to certify this event. The problem for traditional quantum mechanics in that case is that the notion of measurement is built into the idea of a quantum state. A pure state stands for a possible measurement whose outcome occurs with probability-one (e.g., measuring in the same basis as the state preparation). At the same time, Wigner, believing the measurement will occur at the prearranged time, would be aware of all this before the experiment starts. Some authors even say that at that time, Wigner \textit{does} update his state to a classical mixture \cite[14]{zukowski2024}. In our view, if Wigner really does believe in the absoluteness of global events, regardless of quantum mechanics, he should not assign an entangled state to begin with, meaning he is not a sincere user of the formalism. On the other hand, if all his beliefs are rolled into his entangled state, and the theory is complete and universally applicable, then it seems as though personalist measurement outcomes emerge as an implication. Wigner's post-measurement state after the lab is opened is just another update in expectation, not categorically distinct from the one he first assigned. If the event was global, and this were reflected in the collapsed state, then one must retroactively read back ignorance into Wigner's assignment during the time at which the measurement was performed. This would, again, suggest there was ``more knowledge" to be had. When we disentangle quantum states from measurement outcomes, we seem to leave the door ajar for other metaphysical commitments to creep in and to undermine assumptions like completeness or universality. At the very least, it is unobvious whether or not these levels of personalism should be treated as separate.\\

It is common to label Wigner as a ``super-observer” in the following sense: when Wigner assigns a joint entangled state to the Alice-lab system, the presumption is that Wigner’s state assignment allows him to contemplate an action on the system that would undermine her agency \cite[38]{fuchs2023}. We can imagine a situation where Wigner has an experimental apparatus that controls all the degrees of freedom inside the laboratory. According to the unitary nature of quantum mechanics, measurements are theoretically reversible (although “for all practical purposes” irreversible). If Wigner had complete control, then even after Alice performs her measurement, Wigner could turn some knobs and reverse the measurement process, including Alice’s memories of the measurement \cite{del_santo2024}.\\

Less dramatically, Wigner's measurement, owing to the entanglement, can access correlations unavailable to the friend once they have already observed an outcome. QBism rejects that Wigner's entangled state alone is a privileged description of events, partly because it takes assigning a pure state to a system too seriously, as though one could put the friend in “suspended animation,” but also because each state assignment reflects actions and consequences relative to an observer. Fuchs, et al. write of the entangled state: ``Wigner’s state superposes all the possible reports from his friend about her own experience, correlated with the corresponding readings of her apparatus” \cite[5]{fuchs2014}. With a ``pure state” Wigner brings to bear all his possible beliefs about the friend's actions in the laboratory. Note that to fully circumvent the notion of a privileged description, it seems one must view Wigner's prediction as concerning his own future experience, i.e., a personal measurement outcome. It does not seem enough to simply say quantum states are personal to evade a privileged description, as Wigner's state could gesture at an underlying state of affairs that would implicate the friend's state (i.e., the entanglement).\\

The QBist response to the Wigner’s Friend thought experiment exhibits QBism’s understanding of quantum mechanics as a whole. The contradictions that arise in other approaches dissolve when we see that quantum states and measurement outcomes are tied to individual agents. From the QBist point of view, the consequence of treating the quantum formalism as personal is that Wigner and Alice are put on ``equal footing,” making them what might be called ``peer observers.”

\section{Intersubjectivity, Reciprocity, and Transcendence}

In this section, we formulate three levels of intersubjectivity. The first level pertains to general language-use. The second is how intersubjectivity distinctly appears in quantum mechanics, where Wigner and friend are inclined to treat each other as quantum systems. Their situated perspectives matter for their predictions. We call this notion of intersubjectivity ``reciprocity," or how observers reciprocally use the quantum formalism. The third level is specific to QBism, which takes on the further commitment to personal measurement outcomes. Not only do we treat each other as quantum systems in the formalism, but we commit to the physical consequences of this idea. If no measurement outcome has occurred from one's perspective, one cannot update one's quantum state. The QBist notion of reciprocity thus extends the generic notion to our image of the transcendence of the world. If we treat someone as a quantum system, then intersubjectively, we are committed to the idea that the consequences of measurements are relative to a given system. Thus, the intersubjectivity of the formalism itself makes natural personal measurement outcomes through the treatment of another person as a quantum system. The use of quantum mechanics to generate an account of intersubjectivity according to that theory touches more deeply on the meaning of objectification (discussed further in section 8). Reciprocity, for quantum mechanics and for QBism, highlights the intersubjectivity of the quantum formalism, but it is essentially the same as the intersubjective function of language more generally insofar as it provides prescriptions for the standards of correctness for ordering and communicating our observations.\\

First, language-use, including all our theoretical and experimental discourses, is inherently intersubjective \cite{cavell1979, cook1965, kripke1982, habermas1998, sellars1991, wittgenstein1953, husserl1989, zahavi1997, zahavi2017}. If language is already intersubjective, then we do not posit a a guarantee to further make our communications intersubjective. ``Intersubjective agreement" is not an additional criterion to consider after one already has a functioning language. The notions of intersubjectivity, intersubjective agreement, and communication are built into the fabric of being a subject, or for QBism, ``an agent." The way intersubjectivity manifests in the quantum formalism is not exceptional.\footnote{The intersubjective character of language-use and implications has been studied extensively in all traditions of philosophy. If it were possible to have a single subject, why language? In phenomenology, the problem is even starker: is a single subject even a coherent model of subjectivity? The question of intersubjectivity and language intertwines deeply with skepticism/solipsism. The dilemma is just this: if solipsism is not coherent, then it does not have to be ``overcome" as a philosophical problem. ``Solving" skepticism conveys the problem is to be taken seriously. Philosophers often set out to undermine the need for a solution. In philosophy of language, one example is Wittgenstein's famous private language argument, which claims a private language is unintelligible (see p. 344-352 in Cavell's \textit{the Claim of Reason} and p. 83-85, 107 in Kripke's \textit{Wittgenstein on Rules and Private Language} \cite{cavell1979, cook1965, kripke1982}). In philosophy of science, particularly philosophers affiliated with pragmatism, a community, not a single scientist, is presupposed for developing and \textit{using} scientific theory \cite{habermas1998, sellars1991, putnam1995}. In phenomenology, intersubjectivity turns out to be presupposed in the analysis of subjective experience \cite{husserl1989, zahavi1997, levinas1961}. Subjectivity, objectivity, and intersubjectivity are mutually reinforcing notions.} Further, portraying an observer as shut in their own subjectivity, in the context of Wigner's friend, ``courts skepticism" \cite[20]{zahavi2019}.\\

Wigner's Friend can leave the impression that intersubjectivity has broken down, as if each observer occupied their own private universe, inviting us to posit a bridge between subjects that we imagine must already exist for everyday language-use, but not in quantum mechanics.\footnote{In the history of philosophy, particularly modern philosophy, philosophers often began studies of epistemology with a picture of a solipsistic subject constructing an ad hoc bridge to penetrate their shell to an exterior world, the most notable example being Descartes. This includes the question, ``how do I know that the object \textit{you see} is the same as the one \textit{I see}?" Posed in this way, it would seem one needs a metaphysical guarantee to bridge interior and exterior, above and beyond the evidence available to a community of persons. As Stanley Cavell wrote, we, in our skeptical moods, look for "the world to provide answers in a way which is independent of our responsibility for \textit{claiming} something to be so" \cite[216]{cavell1979}.} When one understands that language is  intersubjective, and that the model of a private individual is not the right way to conceptualize subjectivity, then one sees that the bridge we are looking for in Wigner's Friend does not exist for ordinary language either. If the notion of a guarantee on agreement is superfluous for intersubjectivity without quantum mechanics, one need not look for one within quantum mechanics. This imagined connection does not contribute to the objectivity of our claims, including for Wigner's Friend.\\

The usual way philosophers rebuff skepticism is by calling into question the intelligibility of a solipsistic subject, that is, a picture of subjectivity where knowledge of oneself precedes the knowledge or awareness of others. Language-use already accomplishes an intersubjective function in its structure, including standards of correctness and meanings mapped onto a world held in common \cite{frisch1999}. Putnam argued that intersubjectivity is not a ``metaphysical worry" but a practical one. The scientific norms for a \textit{single inquirer} cannot contain a concept of ``truth," which only makes sense when paired with an external standard \cite[70]{putnam1995}. This clues us into the fact that any formalism, including a single-user one, is already intersubjective by construction. If we had to certify that our individual uses of the formalism converge, then individually we must have a definable notion of truth \textit{from the perspective of a single scientist}, which Putnam says prioritizes a private notion of truth over one for an intersubjective community. The intersubjective norm for the quantum formalism instructs an observer how to assign their quantum states for open/closed systems. It does not, in its standard formulation, prescribe that individually their states are privileged or correspond to ``true" reality.\\

An operational formalism articulates a notion of intersubjective truth relative to the theory. This standard of truth gets shared among practitioners when they are trained on the formalism, and that is what, in practice, specifies the possible convergence or divergence of observations. A theory, in other words, may provide guidance regarding the ways in which observers diverge in their observations, such as one that advises us to note the positionality of different observers. The stipulation of meta-criteria would in fact divert us from this conclusion.\\

This lesson from philosophy is instructive for Wigner's Friend. The temptation is to regard the isolation or possible contradiction among participants to mean that intersubjectivity has somehow broken down, and only certain interpretations recover a notion of communication or agreement. However, Wigner's Friend, as we understand it, is not a challenge to the intersubjectivity of experimental results. Rather, a perennial philosophical concern about solipsistic subjects needing epistemic backup for their agreements may have wormed its way into quantum foundations. A clue this has happened is that the problem has the same integrity in classical physics, too (and the same lack of resolution, where it may not be possible to satisfy the skeptic on their own terms). Indeed, there is nothing special about the reporting of measurement outcomes in quantum physics compared to classical physics. Here we are implicitly suggesting that observer-independent facts are not really the sort of facts physicists are accustomed to talking about. The term ``fact" undergoes a sleight of hand when we shift from discussing quantum states and measurement outcomes to asking about a necessary connection among our observations. The former is a fact for CER; the latter is something more akin to an alleged a priori condition for the possibility of science. The connection we are looking for, however, is already made when we use the theory. The evidentiary link is between the definition of a quantum state and a corresponding measurement. If this connection were absent, we would be right to be concerned, as the formalism would not have operational meaning.\\

The quantum formalism seems to prescribe its own particular notion of intersubjectivity which is operative for making objective claims.\footnote{To contrast, philosophers of science, especially in the first half of the twentieth century, styled their activities as ``rational reconstruction," or the construction of epistemology for science, usually by way of a so-called ``theory of truth." See \cite[2]{crease2023} and \cite[10-11]{fine1986} for discussions of this philosophical enterprise. This conception of philosophy of science has shifted over the years to the articulation of objectivity from the vantage point of existing physics, especially utilizing the history of science (see \cite{Daston2007, maudlin2007})}. Note this is not a consequence of ``perspectivalism" but rather a specific theoretical prescription given by quantum mechanics on how to order our observations.\footnote{It is worth flagging the concern that the meta-assertion that facts \textit{must have a point of view to be considered ``facts"} falls into the same trap that we require convergence guarantees for our agreements independently of CER.} In Wigner's Friend, an observer-independent fact corresponds to a ``nowhere point of view." Such a perspective ought to correspond with its own state. If this is a quantum state, it carries the force of real experimental predictions, practical consequences for Wigner and friend. If it is not a quantum state, then one is effectively saying that Wigner's information is incomplete when the friend observes an outcome, which has broader implications for quantum mechanics. Claims for or against observer-independent facts must take into account this putative state assignment. Nevertheless, identifying objectivity with a kind of ``nowhere" point of view is an inherited concept from the history of philosophy, against which scientific theories themselves are often evaluated. The existence of observer-independent facts seems to follow the same pattern. By questioning the relevance of observer-independent facts for CER, and by interpreting quantum mechanics without this notion, one shows how a theory without this feature is not a threat to objectivity or intersubjectivity.\\

With this in mind, it is more natural to frame intersubjectivity in quantum mechanics as ``reciprocity." We reciprocate each other's use of the quantum formalism; Wigner and friend treat one another as quantum systems. From Wigner’s perspective, the friend is a physical system, and from the friend’s perspective, Wigner is also a physical system. The reciprocity norm accounts for situations where agreement about how to use the formalism implies possible disagreement about states or measurement outcomes. That is, the situatedness of observers is physically significant.\\

In terms of reasoning, we suggest the following hierarchy is at work. The formalism is intersubjective and universal, which as many have noted, encourages us to treat others as quantum systems \cite{schmid2024, debrota2020a, brukner2015, adlam2024, schmid2025}. A generic reciprocity norm itself does not single out a quantum mechanical interpretation. The QBist form of reciprocity, in a sense, extends the broader intersubjective principle. Once one holds that the prediction and consequences of measurements are relative to quantum systems, then one would be inclined to see measurement outcomes as personal when the formalism is applied to another person. When we strengthen the reciprocity norm with the claim that measurement outcomes are personal, we are led to the notion that their predictions are independent (to be discussed further in section 8). Each observer can contemplate what the other observer would report if asked about what they saw, but the friend does not have to use Wigner's prediction to make hers correct. We can trade observational reports, but we are in no way inclined to doubt the integrity of another's predictions. The key to the QBist form of reciprocity is to avoid ``stepping outside” one’s perspective and insisting on a global measurement outcome.\\

As we said, we do not think anything particularly extraordinary about intersubjective communication is found in Wigner's Friend. It is often pointed out that when we stick to the first person, we find no contradictions within our situated perspectives \cite{pienaar2025}. Only when we compare predictions do incongruities emerge; certain joint probability distributions cannot be composed. Despite all the variations, however, the salient feature of the thought experiment is still the divergence of measurement records among observers.\\

Here it is often thought that personal measurement outcomes and the reversibility of measurement are the substantive philosophical issues at stake.\footnote{See ``Whose Knowledge?" by David Mermin (2001) for an dramatic technical example of when the ``personal" matters.} Both call into question, in distinct ways, the objectivity of measurement events. According to Del Santo, et al., “The significance of the new no-go theorems lies precisely in the point that they question the objectivity of observable facts which observers perceive, be aware of or may have knowledge about, and not the counterfactual properties of simple quantum systems, which are the subject of Bell’s theorem” \cite[1-2]{del_santo2024}. However, individual measurement events, personal or not, whether we agree on them or not, have not been the gold standard of objectivity in quantum mechanics for a long time.\footnote{See for example, Schr{\"o}dinger's discussion of how the objectivity of the quantum state displaces the objectivity of empirical facts \cite[143-146]{schrodinger2014}.} In this regard, the QBist emphasis on the \textit{connection} between related measurement schemes, rather than the individual outcomes, is a healthy strategy to press the argument toward the ontological significance of other aspects of the quantum formalism, such as the Born Rule.\\

In our view, what should be seen as more provocative is that Wigner's Friend discloses how much intersubjectivity, and consequently our agreements, rely on a constant hotbed of interactivity. Reciprocity, for agents, is peer-based interaction. When this is taken away, the transcendence of the world remains, but acquires an altogether different character. The world no longer looks like a summation of a state of affairs or a plenum of substance. We are given over to an altogether different picture of what our experience of an outside world is like.\\

According to QBism, transcendence is a perpetual ``moreness" that overflows our observations. This is nothing short of an essential commentary on realism.\footnote{QBists and commentators have argued that reciprocity in Wigner’s Friend amounts to a kind of “Copernican Principle.” Such a principle generally refers to a ``decentering” of subjectivity – the equivalence of all agents in their use of the formalism for making predictions on the world \cite{fuchs2023,cavalcanti2021, fuchs2010}.} Wigner's entangled state saturates his predictions, but nature can still do otherwise. In the QBist argumentation, one arrives at this picture of realism because the quantum formalism is not ``descriptive” in the following sense: quantum states are not predicates of systems. Because states are expressions of expectation, one concludes that quantum mechanics is not everything there is, either. Fuchs writes: 
\begin{quote}
    We believe in a world external to ourselves precisely because we find ourselves getting unpredictable kicks (from the world) all the time. If we could predict everything to the final T as Laplace had wanted us to, it seems to me, we might as well be living a dream. To maybe put it in an overly poetic and not completely accurate way, the reality of the world is not in what we capture with our theories, but rather in all the stuff we don’t \cite[8]{fuchs2016}. 
\end{quote}
At first, the transcendence of the world seems like a black box, where the kicks are nature's multifarious responses to our measurements. The black box picture is altogether misleading, however, as it gives the impression that the transcendence of the world has withdrawn. All one would have, then, is the opacity of the quantum interface, the real world tucked away. However, the opposite has occurred. Our contact with the world has intensified. In the history of philosophy, a profound lack of imagination led us to regard nature as veiled because the only activity we could think of as truth-seeking is ``peeling it back" to see what is underneath. The impression of depth bandies with the impression that the veil \textit{itself} is inert and uninteresting. Of course, what we were writing off as a ``veil" could have been \textit{nature} for all we know, but we did not have the tools, or the patience, to prove to ourselves that the seemingly dull surface is as rich and as animate, arguably more so, than what was thought to be kept behind the veil. The search for truth, defined as uncovering and discarding the surface, passes over the target of the search. To embellish the metaphor one more step, physical interactions are not exactly layers we peel back but reflect \textit{density} in the phenomena. Transcendence, ``underlying reality," is concentrated and elaborated in the texture of the interaction. Unlike the black box, we do not get a \textit{satisfactory} description from the disappointment of trying to pry it open, but through the kicks. A measurement scheme poses a question to nature; nature responds. When one has a constant sense of interacting with the world, then each individual kick is unremarkable. Measurement is unremarkable, and a sense of continuity persists. Only in a circumstance where sharing a world is delayed do we note the kick’s absence. In this sense, the isolation of Wigner's Friend is an exception that reveals the norm of reciprocity. When systems are interacting all the time, the QBist tenets of Bayesian probability, personal measurement outcomes, and an adherence to the first person do not seem urgent. Yet Fuchs says, when Wigner and friend are apart, ``the rest of the story is deep inside each agent’s private mesh of experiences, with those having no necessary connection to anything else” \cite[11]{fuchs2016}.\\

QBism prefers to define measurements for single systems, where they are often glossed as ``moments of creation." Nature can surprise us according to QBism. However, there is nothing obvious about the radical transcendence of the world that would imply personal measurement outcomes. Creation could have been \textit{gleich Alles zusammen}. If there is a tiebreaker to be found in Wigner's Friend, it rests with the QBist reciprocity norm. As is generally agreed upon, Wigner cannot drop his entangled state until he opens the lab. However, Wigner may still hold that his state is a privileged description of reality. To secure the claim to personal measurement outcomes, two further conditions for quantum states seem necessary: quantum states are not predicates of systems and also not states of ignorance about global states of affairs. If so, then Wigner cannot undermine the friend's measurement. This move legitimizes the efficacy of the friend's outcome even when an outcome has not occurred for Wigner.\\

Now, despite their isolation, Wigner and friend are still in an inherently intersubjective situation. It is true that intersubjectivity could not arise if the inner life of an agent were confined to a bubble; in fact there would be no coherent notion of an agent. A truly solipsistic subject is no subject at all (they cannot even raise the question of their own solipsism!). Our experience of the world is not bubbly, where the interior is transparent self-knowledge, and the exterior is opaque reality. We carry both senses of interiority and exteriority from our situated perspective. Transcendence must be understood as the actual and potential arrival of novel experience, not as effecting a permanent division between the interiority of the subject and the exteriority of the world. In Wigner’s Friend, one still makes inferences about third-person objectivities, which likely speaks to the resilience of objectivity in physics and the fact that Wigner’s Friend is not a story about challenging objectivity. After the lab is opened, Wigner and friend communicate an experimental result that they intend to be a shared belief about the world. Before this, Wigner includes many beliefs about the friend in his pure state assignment, which is necessary for him to be on equal footing with her as another agent. However, we cannot understand or plausibly hold reciprocity if, from Wigner’s point of view when he assigns a pure state, he treats the friend as a “projection” of his mind. This touches upon the importance of the question regarding the extent to which Wigner and friend can objectify each other as quantum systems, to which we return in section 8.

\section{London and Bauer and Intersubjectivity}

Let’s now contrast the QBist account with London and Bauer's story about intersubjectivity. Their account of quantum measurements comes from the book \textit{The Theory of Observation in Quantum Mechanics} (1982). French has argued convincingly that Eugene Wigner misinterpreted the ``faculty of introspection” in the LB account. LB talk about the ``essential role” of an observer when taking measurements, but that role concerns an observer recording a measurement outcome \cite[251]{london1982}. Nowhere in their book do they suggest the observer plays a causal role. On the contrary, an observer’s quantum state describes a probability distribution for a given system. Quantum mechanics is a linear theory – the sum of two quantum states is also a quantum state. This allows the observer to write a “global wave function” representing states for the observer, the apparatus, and the quantum system \cite[251]{london1982}. However, this quantum state assignment does not reflect on its own a radically different physical situation. LB say that with respect to objectivity, for measuring an object, the combined system of observer-apparatus-object is not much different for an observer than the apparatus-object system (and talking about the movability of Heisenberg cuts works just as well here). In their view, writing down a pure state for a system does not mean something special for the object itself. An essential part in the argument is that the measurement process is not a “mysterious interaction” between an apparatus and an object \cite[252]{london1982}. We take this to mean that LB regard the “ontic” interpretation of quantum states as a misconception. This rules out for them ``objective wave collapse” no matter what we specify as its cause (an interaction, an observer, decoherence, etc.). Rather measurement results in an increase in knowledge:
\begin{quote}
    It is precisely this increase of knowledge, acquired by observation, that gives the observer the right to choose among the different components of the mixture predicted by theory, to reject those which are not observed, and to attribute thenceforth to the object a new wave function, that of the pure case which he has found \cite[251]{london1982}.
\end{quote}
The measurement event replaces the statistics of the global wave function with observed outcomes. Two things arise from this event: an observer and an observed, which we usually abbreviate as ``reporting the result.” The act of observation allows one to ``\textit{set up a new objectivity}” and attribute a new state to the object (original emphasis) \cite[252]{london1982}. LB are assuming the definitive measurement outcome is the objective part of quantum theory (as opposed to the “superposition” itself). However, a measurement outcome is not something we naively or passively record but select from what they call ``a set of ‘potential’ probability distributions or statistics” \cite[259]{london1982}. LB warn against abstracting away details from the measurement context since it becomes meaningless to talk about ``measurable properties as realized” \cite[259]{london1982}. We interpret their advice to mean that one can only talk about a specific set of physical properties of a system in proximity to the relevant measurement of that system.\\

LB also do not think that quantum mechanics radically changes our intuitions about intersubjective agreement. However, their definition is still specific to an experimental context. Intersubjective agreement, in the broadest sense, means that two observers who use the same apparatus will make the same observations. They call the ``coupling” between an observer and the apparatus a ``macroscopic action,” not a quantum one. In other words, writing down a quantum state for an observer and an instrument refers to an action taken on those objects – such as making predictions for another experimenter inspecting the apparatus. One can predict that another observer interacting with the apparatus will report the same outcomes as oneself. While we cannot ignore the ``scrutiny” of an observer when making an observation, since the measurement outcome is an increase in knowledge, we can ignore an observer when examining the instrument itself \cite[258]{london1982}. Therefore, nothing in quantum mechanics prevents two observers from looking at the same instrument before an experiment, making the same predictions, and recording the same measurement outcome afterwards.\\

LB end their account by saying that while objectivity is just as valid in quantum mechanics, the meaning of that objectivity is up for reconsideration. Here they cite Husserl, Cassirer, and phenomenology itself as paths to investigate this change \cite[259]{london1982}.\\

According to French, LB were talking about a kind of correlation that precedes the subject-object distinction. Equally vital is the creative aspect of objectification. A measurement outcome is an outcome for somebody, but we can only make sense of an object as a grasping, a delimiting, or an isolating-in-reflection. The various ways that we “carve up” phenomena are recipes for different objectivities. At the extreme, we may even regard our methods for describing these substrates as new ontologies. These ontologies may then become reified in our imagination, transformed into static idealities, and read back into nature as what we originally found there (spurring Husserl’s critique of ``surreptitious substitution” or Whitehead’s ``fallacy of misplaced concreteness”).\footnote{While similar, these critiques are distinct. Whitehead's argument targets the ontology of abstraction directly \cite[51]{Whitehead1967}. Husserl's argument concerns forgetting or passing over the meaning of our theoretical notions and thus misunderstanding their ontology \cite[48]{husserl1970}.} French states clearly that the reflective action LB identified is not specific to quantum mechanics: ``The issue as to whether there is any more ‘creativity’ - understood in its typical, non-phenomenological sense - in quantum situations as compared with classical ones is irrelevant. In both cases, it is not a matter of whether the observer feels more creative when making an observation in a quantum context or not, since from the phenomenological perspective, the very act of objectification is a creative act” \cite[151-152]{french2023}.\\

The interpretation of measurement in LB's book pays attention to how we group different systems. Intersubjective validity in their account partly depends on, for example, a state for the observer-instrument system being different from one for the observer-instrument-object system. This inclines their analysis toward an ``epistemic” view of quantum states, since one’s state assignment changes depending on what objects are included in the system. The ``cut” between observer-observed is not defined by underlying properties of the system. However, LB refer to an observation as an increase in knowledge, but they do not answer the question of whose knowledge or whose state. If a state is a ``state of knowledge,” it makes sense to ask whose knowledge it is – the experimenters? LB sometimes speak of being ``in” a state \cite[259]{london1982}, which seems inconsistent with the epistemic view. At best, their position on the status of quatum states is ambiguous. It is unclear whether or not LB regard quantum states as being about an observer's knowledge or about the the world.

\section{Critiques of Intersubjectivity in QBism}

We are now in a better position to address French’s concerns about intersubjectivity. French writes of QBism: 
\begin{quote}
    As far as Wigner is concerned, when it comes to his interactions with his Friend, or the system, or both, he should make any decisions regarding the consequences of these interactions according to the prescription given by quantum theory. This prescription will be different as far as the Friend is concerned and as long as we appreciate that difference, the QBist maintains, there is no conflict...Nevertheless, there remains the worry that the central issue of establishing intersubjectivity here has not been fully addressed \cite[194]{french2023}.
\end{quote}
French is correct to imply that reciprocity \textit{does not occur} because of any presupposed intersubjective agreement about their states. Wigner and friend cannot override each other’s use of the formalism so long as they used it correctly. Other situations in quantum information tell a similar story. If Alice ascribes a quantum state to a system, then Wigner cannot clone or copy that state. To figure out what her original state was, he has to ask Alice about her beliefs of that system. Elsewhere, French says that Wigner and friend must agree on their state assignment: 
\begin{quote}
    However, an appeal to empathy is not going to be enough to ensure a common framework of objectivity in this case – what we need is some assurance that not only will the observer and her friend agree that there is an object ``there”, but that they agree as to its \textit{state}, as given by the measurement \cite[175]{french2023}.
\end{quote}
Our question for French is, whose measurement (also, whose state)? Wigner’s or Alice’s? French’s wording seems to presuppose the absoluteness assumption. When we associate the ``state” with the object instead of Wigner and Alice, we are already ``loading” the argument toward hidden variables. French leverages the LB account to argue that quantum mechanics belongs to a ``theory of knowledge” \textit{for observers}, so it is puzzling why he sometimes attributes a state to ``the system” instead of Wigner and Alice.\\

The term ``state” often floats pretty freely in Wigner’s Friend. For clarity, our perspective is that if one grants that Wigner and Alice still have valid quantum states at the intermediate time when someone performs a measurement but before the lab is opened, then one is not imposing a presupposed agreement on their post-measurement states. If one does expect that the conversation between Wigner and Alice reveals an objective event, then one should object to the apparent contradiction in states at the intermediate time after Alice takes a measurement but before Wigner asks about it. Someone ought to update their state after the measurement if an outcome has objective significance.\\

More importantly, without further specifying whose measurement we are concerned about, French could be mixing apples with oranges. His account of intersubjectivity in LB clearly refers to \textit{two observers inspecting the same apparatus and using the formalism in the same way}. In the case of our thought experiment, that would mean \textit{two Wigners or two friends, not Wigner and friend}. Their position in the lab and their access to the instrument matters crucially when making predictions. We thus do not see anything in the LB account that would demand that Wigner and friend agree on their post-measurement states. If we rehearse the standard telling of Wigner’s Friend from LB’s perspective, only the friend could write a quantum state that singles out the apparatus and the object, while Wigner would still have to write a joint entangled state for the entire system. When they compare their results after the lab is opened, nothing in LB entails that their post-measurement states be the same (which either means they have the same problem that French says QBism has, or this is not really a problem at all). Nothing in QBism prevents Wigner and friend from agreeing that an object is there, and nothing in QBism prevents agreement on updating their state assignments after they report to each other. The QBist account rejects the ``necessity” of the situation. What we believe LB are referring to is a situation where another observer competent in quantum theory appears aside from Wigner and \textit{also} writes down a joint entangled state for the friend-lab system. Wigner may further imagine that if he and the friend swapped places, he would use the formalism in a comparable manner. In other words, whatever else it may entail, intersubjectivity in London and Bauer does not necessitate agreement.\\

Emily Adlam has offered a similar critique as French \cite{adlam2022, adlamrov2022}. We address Adlam's perspective because it gives us a chance to distinguish between the concepts of absoluteness and tracking in Wigner's Friend. According to Adlam, QBism struggles to satisfy a criterion of shared facts: “If Bob measures Alice to ‘check the reading’ of a pointer variable, the value he finds is necessarily equal to the value that Alice recorded in her earlier measurement of S” \cite[5]{adlam2022}. Adlam is concerned with what assurance we get that, after the measurement, they agree when they talk to each other. First, we all seem to be in agreement that ``talking to Alice" counts as its own measurement. If we apply a unitary to a system coupled to an ancilla and then project onto the ancilla, we get the desired measurement outcome. But clearly there is a second measurement to check if \textit{another} system got the same outcome. Shared facts is then about consistency. If every communication is a new measurement, then the issue concerns tracking the identity of measurement outcomes through time.\\

Revisiting arguments in section 2 and 4, this concern about shared facts is not specific to quantum mechanics, nor does the sharing of facts pose any practical problem for CER in any context.\footnote{See also \cite{rovelli2024} for the argument that reading metaphysical preconditions into the possibility of doing science undermines the notion of scientific evidence and how our understanding of the world changes over time.} Adlam associates shared facts with ``observer-independent" ones \cite[5]{adlam2022}. Again, our reading is that the notion of ``fact" has shifted from CER to a metaphysical desideratum, which no longer has a clear interpretation. Now, if we drop the term ``necessarily," whose rhetorical force is altogether ambiguous, the shared facts criterion becomes mundane. However, we do not register Adlam's criterion as a worry over shared facts or absoluteness, but rather about the tracking assumption, or the idea that participants can track each other's measurement outcomes reliably. In an extended Wigner's Friend setup, it is assumed that each participant tracks the measurement outcome (Alice tracks Bob's outcome, who tracks Charlie's, and so on). The tracking assumption allows each phase to be treated as part of a single measurement process. If a reverse unitary is applied to someone's measurement, then the possibility of contradictory measurement outcomes is carried forward in Bob's memory. Without tracking, all the runs have to be treated as fundamentally different measurements.\\

Another question concerns whether or not absoluteness implies tracking \cite{schmid2024, zwirn2025}. In our view, they are indeed independent, in which case insisting on absoluteness does not satisfy any guarantee on fact sharing. The basic reason is that an event is not in and of itself a fact, the latter which conveys epistemic content.\footnote{Per our theme of the history of philosophy, the existence of a class of facts which tender immediately legible content of nature, that one could ``read off" the object, deserves to be called ``analytic." The problems with analytic statements or de facto empirical truths are well known, and were abandoned in philosophy of language long ago. Suturing absoluteness to tracking, as if the event itself gave the fact, strikes us as a strange proposition, whose aspiration is not altogether different than the analytic report. One does not find epistemic fruit, so to speak, hanging on the vine.} One can easily imagine a world with objective events, but whose intelligibility is scrambled, such that identifiable facts could not be extracted from them. Rather the question of tracking concerns identity in time (akin to Heraclitus' adage that one does not step in the same river twice). Absoluteness is neither a necessary nor sufficient condition for tracking, nor is the question of tracking unique to physics. One does not need quantum mechanics to ask, ``how do I know that the entity you saw is the same one that I saw? Or the one I see now will be the same one that I measure in the future?" Here Alice and Wigner may both believe that a measurement outcome really occurred in the way their shared report suggests, but our point is that if we really doubt this idea, we will appeal to concrete details in the experiment to explain the success or failure of the corroboration. Tracking would be a fundamental concern for shared facts in the way suggested by Adlam if a measurement outcome was taken as a primitive concept, something that resists physical interpretation. Then one could object precisely that there is no pre-theoretical way of looking at nature which simply dictates ``what is" in a way that everyone ought to recognize. One would be seriously conflating the measurement event with its presentation in a factual claim, i.e., how the event is modeled in a given theory. However, this thing we call a ``measurement outcome" already has a robust description in quantum measurement theory that any user of the quantum formalism would recognize. As such, we are unconvinced the shared facts criterion is an obstacle for any of the interpretations.\\

Following the argument upstream, again, we tend to think that the difficulties of Adlam's and French's criticisms stem from overlooking an intersubjective function that the quantum formalism already has. We are by no means taking a definitive stance on what objectivity in quantum mechanics fundamentally means (which is an open question), but the relevant standards of correctness for doing experiments are already in place. We realize there is a certain amount of irony in this reversal. ``Realist" interpretations of quantum mechanics are correct to be unconcerned with intersubjective agreement, as intersubjectivity is presupposed, whereas solipsistic accounts are inclined to search for metaphysical criteria to prevent so-called ``slips" into the solipsism already assumed as a premise of the search. A sign of this worry in one's thinking is the notion that reality itself requires conceptual supports or scaffolding, for which absoluteness is an example. French obviously shows sympathy to our points above when he quotes LB: ``the discussion of the formalism taught us that the apparent philosophical point of departure of the theory, the idea of an observable world, totally independent of the observer, was a vacuous idea” \cite[142]{french2023}. What may be less obvious, or more contentious, is how the notion of ascribing supports to reality may be rooted in a defective notion of subjectivity.\\

We list three more reasons that we may feel like intersubjective tissue is missing in Wigner’s Friend:

\begin{quote}
    1) An adherence to subject-object dualism makes us feel like we are intrinsically cut off from the world and each other. If measurement outcomes are personal, and personal is interpreted as ``cut off,” then we would like to invent some connective tissue for pulling them together. The olive branch that we communicate with each other would be dissatisfying since that communication \textit{for us} would be just another personal measurement outcome.\\

     2) There is no categorical difference between inspecting the device and talking to Alice. To appreciate this is to see that consulting an impersonal object is just as efficacious as checking the coherence of our observations. To ask for more is to bypass or minimize the relevance of all the actual physical reasons we might give to call into question the efficacy of an outcome (e.g. if Alice has short-term memory loss or the ``click" of the detector was off).\\

     3) We feel as though Wigner’s entangled state assignment somehow does not include his expectations for eventually learning about Alice’s measurement outcome and therefore some extra piece of justification is necessary for why he learned what he learned after the experiment. However, if this were true, then we would be saying that Wigner was missing crucial information, and his quantum state should reflect this.
\end{quote}
1) and 2) are more or less self-explanatory. The crux of disagreement seems to crop up with 3), to which we now turn. 

\section{A QBist Response to French's and Adlam's Critiques}

Now we review, in an introductory way, a QBist argument that asking for a guarantee on agreement is akin to asking for a hidden-variable theory. Our basic problem is that a convergence guarantee for Wigner's Friend seems to indicate the existence of a non-disturbing intermediate measurement to preview Alice's measurement outcome, but QBism suggests this intermediate measurement makes a difference in how we structure our probabilities. The usual way they tell this story involves interpreting the Born Rule as a deviation from the classical law of total probability, where the quantum case is irreducible to the classical one.\\

What could guarantee Wigner and friend to agree? One could argue, it should mean something concrete and practical if agreement is inevitable.\footnote{Several commentators pointed out to us the similarity between this demand and Spekkens' Leibniz methodological principle \cite{spekkens2019}.} With this line of argument, one would be assuming that one cannot recognize a description given from one's own perspective as correct while simultaneously demanding ontic inevitability from a perspective given from nowhere. In other words, do we disbelieve Wigner’s assignment of an entangled state? From an imagined bird’s eye view, this perspective would naturally regard Wigner’s entangled state as incomplete information, since the bird’s eye already ``knows” what is happening in the lab with Alice. From this perspective, associating a state with the system after the lab is opened would imply a fact of the matter appeared for Alice before Wigner inspected the lab. This reasoning implies the existence of another ``better” state that Wigner did not assign. Of course, this state could really be a part of Wigner’s entangled state too, as Wigner may imagine that he really does know what is going on inside the lab. Is there a mismatch between Wigner’s state and what he really believes? If so, a measurement for Wigner should be forthcoming. If one can assure agreement, it should be possible for Wigner to conceive of a measurement strategy and preview the measurement outcome before the lab is opened. This would allow him to circumvent a report from Alice. Yet, what could this be except a hidden variable? Consider the following measurement scenarios.\\

First, Wigner could actually check what Alice saw after she had performed a measurement but before she tells him. Wigner’s probabilities for hearing Alice’s report \textit{E} given that he sneaks a peek into the lab and finds \textit{R} are different from talking to Alice directly. There is nothing uniquely quantum-mechanical about the difference in the probability assignments $P(E)$ and $P(E|R)$. If I make a prediction about tomorrow's weather $P(E)$, it is different from the probabilities I would assign by checking the weather report first $P(E|R)$. In this way, it is not surprising that checking the aperture in a double slit experiment before looking at the viewing screen gives different probabilities than just looking at the viewing screen. Probability theory already tells us there should be a difference.\\

Second, without intending to perform a measurement, Wigner can imagine himself checking what Alice saw before she tells him. He can even feel certain that Alice will report the same observations he would have seen. In any given experiment, we can always make a hypothetical prediction for a measurement that we do not actually do. We make these predictions all the time. In the case of the double slit experiment, we could imagine putting a detector in the path of the electron beam to check whether an electron goes through one slit or the other. An agent can consider a direct measurement \textit{E} or hypothetically putting the system through a measurement \textit{R} first before acting on the system. When this first measurement \textit{R} is not actually performed, the probability $P(E|R)$ is not defined. Can we intuitively assume that the counterfactual $P(E|R)$ will be the same as given in the paragraph above?\\

Now, one is required to use quantum mechanics to answer the question, ``does it matter if one intends to do an intermediate measurement or not?” In most classical situations, it does not matter. If the measurement $R$ simply reveals an outcome, then the probabilities $P(E|R)$ should be the same regardless if $R$ is performed. However, in the double slit experiment, an intermediate measurement on the aperture destroys the interference pattern. In Wigner’s Friend, Wigner assigns an entangled state to the friend-lab system. The hypothetical intermediate measurement matters because if Wigner decides to open the lab, he would then not assign an entangled state.\\

The QBists developed an analogy to the double slit experiment using the concept of a reference device. One can uniquely write a quantum state as a probability distribution over the outcomes of a reference measurement as long as the measurement is informationally complete. For one experiment, we perform a direct measurement $E$ on a quantum system whose outcomes occur with probability $Q(E_j)$. In another we perform an intermediate reference measurement $R$ first. The law of total probability by itself relates the probabilities $P(E)$, $P(E|R)$, and $P(R)$ in the experiment where one performs $R$
\begin{equation}
    P(E_j) = \sum_i{P(E_j|R_i)P(R_i)}.\\
\end{equation}
This is simply a statement of coherence among different probability distributions. Here probability theory alone does not tell us how to relate the same event $E$ under two exclusive sets of experimental conditions.\\

In the classical case, we assume there exists a ``non-disturbing" intermediate measurement $R$. Such a reference measurement would simply record the physical conditions of the system, such as its phase space configuration. It is like taking a photograph with an arbitrarily rich resolution, one sufficiently fine-grained that every other measurement can be obtained by coarse graining its outcomes. If a measurement is truly non-invasive, then performing $R$ ought to leave the overall experiment unchanged (assuming we ignore or discard the observed outcomes of $R$). One sees from Eq. 7.1 that averaging over the possible outcomes of $R$ is equivalent to not performing it at all. The protocol with $R$ would be the same as the protocol without it, so that the probability $Q(E_j)$ would equal $P(E_j)$. One could thus express $Q(E_j)$ as a mixture of conditional probabilities given the outcomes of $R$.\\

In the quantum case, the two experiments with probabilities $Q(E_j)$ and $P(E_j)$ are not the same due to the nature of the reference device \cite{debrota2020b, fuchs2023, debrota2021}. One calculates the probability $Q(E_j)$ using the Born Rule by modifying the probabilities on the right side (including a term called the “Born Matrix”) \cite[23]{fuchs2023}. Using vectorized notation, the distinction between classical and quantum is apparent in the altered law of total probability:
\begin{equation}
    Q(E) = {P(E|R)\Phi P(R)}.\\
\end{equation}
The new expression – a modified version of the Born Rule - does not reduce to the classical law of total probability ($\Phi \neq I$). Because the probabilities $Q(E_j) \neq P(E_j)$, the experiment involving the reference measurement $R$ is distinguishable from the experiment with the direct measurement $E.$ The Born Rule now concerns our ability to distinguish two experiments that cannot be performed simultaneously (reminiscent of Bohr’s complementarity). The conceptual picture of reference devices provides key insights into the Born Rule. What changes in quantum mechanics is how the probabilities are meshed together across two different experiments. It is no longer a statement about the probabilities of a single measurement but rather instructs how these tests fit together in probability theory. The QBists see the Born Rule as a normative constraint on our probabilities for \textit{R} and \textit{E}.\\

If Wigner could devise a test that would predetermine their agreement, then he should be able to perform an intermediate measurement on the friend-lab system that he could theoretically coarse grain to preview Alice’s outcome. In other words, a convergence guarantee is likely akin to an intermediate measurement. One can also foresee the difficulty of this move when Wigner assigns his state to the friend-lab system, since measuring the entangled state destroys the entanglement. What the reference device concept clarifies is the operational consequence of this idea: the argument is not simply about disturbance, but the distinguishability of the two experiments.\footnote{More specifically, the distinguishability of the two protocols has more to do with the shape of quantum state space compared to its classical counterpart. The quantum law of total probability is a deformation of the classical law of total probability, where the space of quantum probabilities is a restricted subset of the space of classical probabilities. Technically, the distinguishability of the two protocols is agnostic about disturbance.} Operationally, no reference measurement exists that Wigner can perform and then coarse grain its outcomes to reproduce the direct action of opening the lab. The nonexistence of such a reference device spells problems for convergence, i.e., that Wigner's and Alice's outcomes agree without communication. Stacey puts it this way: 
\begin{quote}
    Two orthogonal quantum states are perfectly distinguishable with respect to some experiment, yet in terms of the reference measurement, they are inevitably overlapping probability distributions. The idea that any two valid probability distributions for the reference measurement must overlap, and that the minimal overlap in fact corresponds to distinguishability with respect to some other test, expresses the fact that quantum probability is not about hidden variables \cite[8]{stacey2023}.
\end{quote}
To predetermine agreement, Wigner could posit a hidden variable that he simply does not know, but there is no corresponding measurement that could reveal it before opening the lab. Nor can Wigner imagine his pure state ``epistemically capturing” this sort of information about the system.\footnote{See also Caves and Fuchs (1996) and Fuchs (2002) on the contents of the quantum state \cite{caves1996, fuchs2002}. Fuchs writes: “The theory prescribes that no matter how much we know about a quantum system — even when we have maximal information about it — there will always be a statistical residue. There will always be questions that we can ask of a system for which we cannot predict the outcomes. In quantum theory, maximal information is simply not complete information” \cite[11]{fuchs2002}.} No measurement scheme to sneak around Alice's report is forthcoming. There is no shortcut to Wigner and friend communicating. One wonders, is it so bad to ask Alice directly?\\

Further, it is worth pointing out that the choice between ``ontic” and ``epistemic” interpretations of quantum states does not by itself resolve the contradiction in Wigner’s Friend \cite{harrigan2010}. Whether one sees quantum states as ``states of the system” or ``states of knowledge,” one’s state can still refer to a state of reality and our incomplete information. In the ontic case, Wigner’s state contradicts Alice’s result the moment she takes a measurement. In the epistemic case, Wigner’s state still conflicts with Alice’s once she gets a result as long as we see Wigner as gesturing at the same state of affairs as Alice. In either case, absoluteness still holds as a hidden variable theory. Brukner has a nice clarification of the politicking surrounding the epistemic-ontic distinction:
\begin{quote}
    [T]he distinction between a realist interpretation of a quantum state that is “psi-ontic” and one that is “psi-epistemic” – which actually is a distinction between two kinds of hidden variable theory – is only relevant to supporters of the first approach. An alternative exists. The quantum state can be seen as a mathematical representation of what the observer has to know in order to calculate probabilities for outcomes of measurements following a specific preparation \cite[5-6]{brukner2015}.
\end{quote}
French is aware that these other possibilities, which evade the dichotomy between the epistemic and the ontic, are not ``instrumentalism.” He writes: ``phenomenologically the wave function is neither ontic nor epistemic insofar as it describes that fundamental correlative relationship which is encapsulated in the slogan, ‘[n]o object without a subject and no subject without an object’” \cite[232]{french2023}. QBism is also neither psi-ontic nor psi-epistemic according to how Brukner defines this distinction, since quantum states are anticipations of an agent’s actions on quantum systems. If one appreciates all this, then it is unclear what would motivate the desire for convergence guarantees.

\section{Objectification as Delimiting Transcendence}

In the preceding sections, we ran the gamut of the features we find unremarkable and the various red herrings in Wigner's Friend (intersubjective agreement, communication, personal measurement outcomes, different quantum states, tracking, and absoluteness). In this section we outline a phenomenological perspective on how QBist reciprocity shifts one's understanding of a quantum system.\\

Reciprocity refers to the reciprocal use of the formalism, or treating each other as quantum systems. When we treat each other as quantum systems, we recognize that the positionality of those systems matters for measurement schemes like Wigner's Friend. An outcome may have occurred relative to one system (Alice) but not another (Wigner). One sees how this reasoning leads to the conclusion that measurements are relative to systems, hence the QBist notion of reciprocity that Wigner and friend are treated as equals.\\

The main objection to reciprocity is that Alice should base her prediction on Wigner's. Baumann and Brukner (2019) showed that Wigner's and friend's measurement outcomes are interdependent if we suppose quantum states correspond with states of affairs. Alice observes an outcome and then computes the probabilities for the superposition basis, $p(+)= p(-) = \frac{1}{2}$. She passes a slip of paper to Wigner with her prediction for the measurement \textit{he will make} on the entire lab. Wigner, of course, assigns a pure state corresponding to the entangled basis, $p(\Phi^+) = 1$, $p(\Phi^-) = 0.$ After the lab is opened, Alice convinces herself that her collapsed state did not take into account the entanglement from Wigner's measurement basis. Arguments along these lines motivate an asymmetry between Wigner and friend, where the friend should base \textit{her} statistics on Wigner's description \cite{frauchiger2018, baumann2019}.\\

The QBists typically reject this reasoning because of possible self-reference. Wigner's prediction for what happens in the lab concerns the friend's actions. If the friend adopts Wigner's state as her own, and Wigner's state is about the friend, then she indirectly is assigning a state to herself. Prima facie, the reason QBists reject self-reference is that there must be a “clear separation between agent and measured system” \cite[10]{debrota2020a}. What exactly is the objection? The friend's state now depends on measurements she has promised to perform. DeBrota et al write: “But since she is a free agent, she has control over the answer to this question. It is up to her whether she replies ‘up,’ ‘down,’ or by sticking her tongue out." \cite[10]{debrota2020a}. Thus the friend's quantum state no longer gives a reliable prediction for her future experience. Not only is it paradoxical to make predictions about one's own actions, in the above case, one is also not using the formalism to calculate one's probabilities. Combining this reasoning with ``no absoluteness" principle and a strict adherence to the formalism, there is no privileged position-taking that would tell us whose facts are better. To see that the positionality of Wigner and friend matters for the quantum formalism is to see that one must stick to their own perspective, i.e, not try to use another's quantum state.\\

How should phenomenologists understand this move? The common practice is to represent subject and object as poles of a so-called ``correlation," where the subject-pole is termed the ``noesis," the object-pole the ``noema." The maxim is: no subject without an object; no object without a subject. We distinguish subject and object as a dipole of one relation, but neither is reducible to each other. This amounts to saying that self-objectification is permissible, but should not be understood as complete \cite[255-256]{husserl2001}. A distinction emerges between subject and subject-as-objectified \cite[84-87]{merleauponty2012}. One may objectify themselves as a quantum system or another as a quantum system according to the theory, but one cannot turn themselves into an object for themselves. The phenomenologist Merleau-Ponty's analysis of the phantom limb is a case study typically thought to expose the ambiguous distinction between our subjective embodiment and thinking of ourselves as merely flesh, as body-object. In other contexts, this delimited description is of limited application. For example, in QBism, one can neither treat the body as a classical measurement device anymore than we can treat an actual measurement device with classical concepts in every situation \cite{pienaar2020}. We should be wary of a misunderstanding: the distinction induced between phenomenon and phenomenon-as-object is a counter to reification, which portrays nature as a ``heap." The distinction does not arise because some phenomena are intrinsically more special than others. We are given over to a distinction between objectifying something as an ``object" according to a specific theory and turning something into an object full-stop, as a piece of pre-theoretical reification. The former recognizes that objectification is itself a theoretical activity. Regarding the latter, the move to call something an ``object among objects" faces a difficult problem, namely, the multifarious character of modeling in physics. This line of thinking - more so than the separation of agent and world - informs QBist reasoning about self-reference.\\

The QBist Jacques Pienaar has nicely refined this point about self-reference in a way that brings the argument into focus:
\begin{quote}
     In QBism nothing prevents an agent from positing a \textit{model} of themselves in order to predict their own actions and behaviours under some hypothetical circumstances removed from their present activity, however, the ``self” which is thereby modeled \textit{is not the self that is doing the modeling}. Attempting to directly include the \textit{model-making self} within the very model that it is making would lead inevitably to logical self-reference paradoxes, making it internally inconsistent \cite[9]{pienaar2025}.
\end{quote}
What matters for quantum self-reference is a distinction between the subject and the subject-model. Generalizing, QBism has cultivated a way of thinking about quantum states that makes the act of objectification explicit.\\

When we objectify others, we face a similar situation. Husserl observes that objectifying ordinary things is entirely different from objectifying another person: ``the apprehension...of whom we are conscious as a real person…\textit{seems to contain a surplus}” \cite[147, original emphasis]{husserl1989}. No doubt Husserl is referring to their inner life. However, their surplus of presence does not have anything to do with the intrinsic ``success” or ``failure” of objectification. His comment does not suggest that a person is a black box, whose inner life presents itself to us as a surplus beyond our knowledge (the platitude ``I cannot know another mind" is the tempting but wrong conclusion to draw about intersubjectivity). Like with nature, we tend to see others as ``veiled" and look past the potent expressiveness of our bodies for the mystery ``behind the face." Husserl's point was more like this: objectification is inherently thematic and selective according to an existing set of theoretical resources. To objectify something is to bring a phenomenon under a specific theme. There is an intentional choice about how to express or theorize the transcendence of another in the same spirit as modeling one's own subjectivity. Husserl expressed the role of interest in the act of objectification: a theme ``designates the object as the substrate and center of a unitary interest” \cite[290]{husserl2001}. Here ``theoretical conceptualization" plays the same role as ``interest" because they each delimit transcendence in a suggestive way. When we represent another person as an object, we are regarding them in a directed, interested manner according to all the relevant notions we can collect when calculating our probabilities for the outcomes of future actions. A ``surplus" will be left over because objectification is scoped by theory, but the nature of the objectification depends entirely on the kind of theorizing we are up to.\footnote{This is a more delicate point than we have space to explore. We are often given the impression that theorizing succeeds by conceptualizing phenomena in a manner that is empirically successful, and therefore ontologically efficacious. Yet the act of conceptualization also shapes the object of inquiry, making it conform to all the things we can say about it, even when clearly there are many more things to say! Call this the ``pessimistic view." The optimistic view is to take on the transcendence of the world itself as an ontological commitment and use that to explain the selective nature of objectification, not for a lack of knowledge, but as an expression of deep purpose. One is warranted to wonder if this is the difference between glasses half-empty or half-full.} To treat another person as an object carries an obvious moral connotation, but for Husserl, objectifying a person as a thing was legitimate within certain bounds: 
\begin{quote}
    I treat a human theoretically as a thing if I do not insert him in the association of persons with reference to which we are subjects of a common surrounding world but instead take him as a mere annex of natural Objects which are mere things and consequently take him as a mere thing himself \cite[200]{husserl1989}. 
\end{quote}
In Husserl's thought, treating another person like an object at first looks identical to treating them like a person \cite[189]{husserl1989}, but only by treating others as persons do we get an objective picture of scientific community. Part of Husserl’s argument pertains to French’s concerns. One can only understand the objectivity of objects after we apprehend the objectivity of others intersubjectively. When we objectify a ``piece” of reality, one has to know that one is creatively carving a subsystem out of a whole. To do that, one has to know that objectification – representing a piece of reality with one's ideas – belongs to something outside oneself. Often the otherness of the Other – another subject – is taken to be the transcendence we need in order to understand this relationship \cite{levinas1961}. That is, knowledge of an external world is conditioned upon the knowledge of persons. If an Other presents as a ``surplus of presence,” then the phenomenon of otherness is essential for knowing that objectification delimits transcendence, which is true for ourselves, persons, and objects. To phrase it another way, it is common to find philosophers, especially phenomenologists and pragmatists, claiming that the ``surplus" of ordinary things is conditioned on recognizing the ``surplus" of others \cite{putnam1995, levinas1961}. Here intersubjectivity does important conceptual work, both for grounding the transcendence of the world and for realizing that objectification is a theoretical activity, not a passive, naive reading of ``what's there." The distinction between a system and our beliefs about a system is maintained even when our expectations for their behavior are certain. Probability-one, in another sense, represents a kind of saturation point for a given set of expectations. Yet due to the thematic character of objectification, it does not efface the ``surplus of presence,” since to call something a ``surplus” at all is to recognize that something as intrinsically transcendent.\\

So much for phenomenology: can the same be said of quantum systems? Whatever QBism wants to do with quantum states, we argue it should be the same for ourselves, others, and objects (part and parcel of the universality assumption). The restriction on self-reference is likely a subclass of how objectification works more generally. If it is problematic to turn an agent into an object, then similar problems likely also hold for any system. Self-reference surfaces the dynamic of objectification, but there is no reason to say it is limited to cases of self-reference only. To write a state for a quantum system is to objectify it as one. There are two senses of objectification worth unpacking. The first is theory-dependence; the second is genuine transcendence, or nature doing what it wants to do.\\

First we do not have a concept of erasing the objectivity of a system, i.e., fully objectifying it. We adopt a thematic stance toward a phenomenon and treat it as the kind of object in the theory. The phenomenology of objectification, from the perspective of QBism, is not only thematizing but action-taking. To identify this system is to ``conceptually carve it out of its background and ask how best to gamble on later, further actions upon it” \cite[20]{debrota2023}. The upshot is that when we use terms like ``system,” we do not mean a generic ``object,” but a conceptualized substrate, the phenomenon brought under the theme.\\

Second, somewhat ironically, removing the stipulation that quantum states are about global events or hidden variables surfaces the transcendence of quantum systems. It shines a spotlight on a sophisticated notion of realism in QBism. What has actually changed is the notion of what a quantum system is. Fuchs has a slogan: “no matter what quantum state assignment is made to a system, the system is more than that” \cite[37]{fuchs2023}. Stacey gives a positive designation to quantum systems in the absence of hidden variables: they exude “vitality” \cite{stacey2023}. If one has convinced oneself that quantum states are not states of systems, but statements about anticipating future experience, then characterizing a quantum system with such a state is also an anticipation. The quantum system itself, as part of nature, lies outside our agency. QBist reciprocity in Wigner's Friend does the most work to expose this commitment to the intrinsic transcendence of the world and of others, as it is another agent who surprises us in Wigner's Friend, not just any quantum system. In any case, these are the sort of thoughts one has under the assumption that the ‘cup of reality really does floweth over.’

\section{Conclusions}

In this article, we explored the QBist notion of reciprocity to address concerns about intersubjectivity. We argued that the quantum formalism is already as intersubjective as anyone could reasonably ask for. We showed that a convergence guarantee implies a measurement scheme inaccessible for both Wigner and friend. Finally, we suggested that guarantees pertaining to intersubjective agreement are rooted in a solipsistic view of subjectivity and altogether ignore the relevant physical reasons we would doubt our observations. Additional demands for intersubjective agreement can be made independently of any theory, which is indicative of a problem with the criterion, not quantum mechanics. Currently, we see worries about intersubjectivity regarding differences in quantum state assignments and personal measurement outcomes as unfounded.\\

When we ladle into the reciprocity norm the thesis that measurement outcomes are personal, we get the view that the formalism makes Wigner and friend ``peer observers." We suggested that reciprocity stimulates further investigations into objectification. The QBist reading of the quantum formalism recognizes phenomena that transcend our agency, and this insight is cashed out for all sorts of entities we designate as ``quantum systems." Objectification, for QBism and phenomenology, is an activity of treating phenomena in the way prescribed by the theory. Reciprocity thus discloses a provocative way of conceptualizing the transcendence of the world, as something \textit{radically} external.\\

Here one of the dangers of saying that QBism requires a ``theory of the agent” specifying its nature is that we risk going behind the sciences completely to dictate how agency ``really" works and thereby stipulate a priori how agency must look in quantum mechanics. QBism’s account of agency surely does not depend on the results of some other debate from another arena. It is more productive for phenomenology and QBism to identify commonalities within their own conceptual and methodological architecture and develop these sites of mutual engagement.\\

On this point, QBism’s ``surplus” of transcendence in its treatment of quantum systems surely throws for a loop Husserl’s claim that objectification cannot accommodate the intrinsic surplus of an object. Husserl implied the kind of objectification that goes on in the theoretical sciences is so exacting and mathematical as to \textit{preclude a full recognition of reality}, including the reality of the person. After all, a theory of knowledge has to represent nature somehow even when we are acutely aware that we are engaged in a particular way of looking at nature to accomplish this task. It is worth bearing in mind that the natural attitude represents a set of ontological commitments we take for granted, whereas the naturalistic attitude is said to refer to the natural sciences. Just as there is no ``'the' natural attitude" that we essentialize into one ontology, there is no ``'the' naturalistic attitude" that essentializes the sciences into one epistemic quality. Husserl was more concerned about the ``forgetfulness” of these attitudes when we slip from one way of thematizing nature into another, prompting us to ``absolutize its world, i.e. nature” according to one uncritical theme of objectivity \cite[193]{husserl1989}. The difficulty a phenomenological framework seeks to overcome, Husserl wrote, is that we ``constantly slip...from one attitude to another” in such a way that important ontological distinctions pass unnoticed \cite[190]{husserl1989}. Nothing precludes a transformation in the self-conception of science that internalizes a characterization of systems as surpluses of qualitative relations instead of a narrower view of what an object is.\\

While much more can be said about the relation between phenomenology and QBism, we do not believe that discussion will lead one to endorse the view that intersubjective agreement about measurement outcomes is a steppingstone to objectivity. Objectivity, whatever it is, will have to be found in the theory. QBism regards quantum states as subjective judgments, but that does not mean the quantum formalism is subjective. Nor does it mean that quantum theory exhausts everything interesting to say about reality. QBist reconstructions and phenomenological approaches to quantum mechanics subvert some ambitions of traditional objectivity and fulfill others. In the position explored in this article, systems exceed their objectification, but such a shift in perspective seems only possible when one gives up the view that quantum states correspond to states of affairs.\\

\noindent Acknowledgments: Gino Elia acknowledges support from the John Templeton Foundation through Grant 62424. The opinions expressed in this publication are those of the authors and do not necessarily reflect the views of the John Templeton Foundation.

\bibliographystyle{unsrt}
\bibliography{wig.bib}

\end{document}